%% file: hetnet.tex
\documentclass{article}
\usepackage{graphicx}
\usepackage{cite}
\usepackage{amsmath}

\begin{document}

\title{Integrated Optimization of Heterogeneous-Network Management and the
Elusive Role of Macrocells}

\author{Raphael~M.~Guedes\\
Jos\'e~F.~de~Rezende\\
Valmir~C.~Barbosa\thanks{Corresponding author (valmir@cos.ufrj.br).}\\
\\
Programa de Engenharia de Sistemas e Computa\c c\~ao, COPPE\\
Universidade Federal do Rio de Janeiro\\
Centro de Tecnologia, Sala H-319\\
21941-914 Rio de Janeiro - RJ, Brazil}

\date{}

\maketitle

\begin{abstract}
We consider heterogeneous wireless networks in the physical interference model
and introduce a new formulation of the mixed-integer nonlinear programming
problem that addresses base-station activation and many-to-many associations
while minimizing power consumption. We also introduce HetNetGA, a genetic
algorithm that can tackle the problem without any approximations. Though
unsuitable for practical deployment, HetNetGA enables the investigation of such
networks' true possibilities. Results for scenarios involving both macrocells
and picocells often align with what is expected, but sometimes are unexpected
and essentially point to the need to better understand the role of macrocells in
helping provide capacity while remaining energetically advantageous.

\end{abstract}

\newpage
\section{Introduction}
\label{int}

In order to meet the rapidly increasing demand for wireless capacity, a current
tenet is that future-generation networks will have to rely on heterogeneity to
grow. That is, every large base station (macrocell) deployed will have to be
accompanied by a number of small base stations (picocells) spread amid users to
improve capacity. The resulting network density will make the problems of
managing interference, meeting capacity demands, and saving power not only more
pressing but also more tightly coupled with one another and consequently harder
to solve. Because of the many trade-offs involved in all decisions related to
tackling these problems, at least some of their key aspects will likely be
handled as a single entity. These include deciding which base stations to be
turned on and which associations to establish between base stations and users.
Of course, such decisions will always have to take into account the way the
available spectrum is handled so that, in the end, the resulting contributions
to network capacity are as strong as possible.

In the last few years, a number of works have addressed, with varying degrees of
detail and success, the formulation and solution of optimization problems
targeting more than one of those aspects concomitantly (cf.\ 
\cite{ku16,zgh16,zjll18,aav18,aazz19,ltlceq20} as representative examples).
While most of them differ significantly from one another, they all share some
important features, including optimization problems of a mixed nature (integer
and continuous) that are handled only after undergoing approximations or having
their feasible sets substantially reduced through the adoption of a limited set
of so-called ``patterns.'' Moreover, it is generally unclear how the
signal-to-interference-plus-noise ratio (SINR) threshold that is typical of the
physical interference model is handled. In some cases, results are downright
irreproducible.

Here we address the problem of minimizing a network's power consumption while
handling the SINR threshold appropriately, determining which base stations to be
turned on as well as associations between base stations and users, and meeting
user demands for capacity. Each base station can be associated with multiple
users, and conversely each user with multiple base stations. To the best of our
knowledge, ours is a complete formulation that stands apart from previous ones
in at least one feature. In particular, our approach is unique in that we handle
the resulting optimization problem, with all its integralities and resulting
nonlinearities, as it comes. Instead of bending its intrinsic combinatorial
nature to fit some optimization method of choice, we leverage the inherently
stochastic, parallel nature of evolutionary methods and use a genetic algorithm
within a simple exploratory methodology. We present results, all reproducible,
in some scenarios.

Some of these results fall smoothly in line with what has become expected of
heterogeneous wireless networks. This includes the effect of increased bit-rate
demands on power consumption as well as on feasibility. Others have been
unexpected, suggesting that the role of macrocells in such networks may be less
clear than generally assumed thus far. In particular, for the network model and
parameters used, we have been unable to unambiguously pinpoint a situation in
which the combined use of macrocells and picocells would achieve feasibility
while using picocells alone would not. That is, we have found no situation in
which the use of macrocells would be energetically advantageous.

\section{Contributions}
\label{con}

Minimizing power consumption while at the same time deciding which base stations
to be turned on, deciding which associations between base stations and users to
establish, and ensuring that user demands for capacity are all met usually
amounts to a daunting problem, full of nonlinearities and often
non-differentiabilities as well. Solving this problem lies at the heart of
heterogeneous-network management, so for practical deployment both network
models and the algorithms to be used must be simplified in order for efficiency
and scalability to be achieved. The downside of such simplifications is that the
operational decisions they lead to may fail to save as much power as possible or
to meet user demands when they could be met. Thus, while reconciling efficiency
and scalability with solution quality is unavoidably fraught with difficult
trade-offs, the need remains for approaches that do not target practical
deployment but rather the detailed study of a network's true properties. In this
letter we contribute one such approach by introducing a complete model (in
Section~\ref{net}) and a formulation of the associated optimization problem that
makes no simplifications (in Section~\ref{opt}, with the appropriate solution
methodology given in Section~\ref{exp}).

Our model is complete in the sense that it incorporates crucial elements omitted
from previous models. Most notable of all is a clear treatment of how SINR
thresholds are handled for proper decoding. This is lacking in previous models
\cite{ku16,zgh16,zjll18,aav18,aazz19,ltlceq20}, which may result in poor
interference coordination. Our model also provides for the determination of
base-station activation (unlike \cite{aav18}) and which users to associate with
which base stations in a many-to-many fashion (unlike \cite{aav18}, where
associations are not considered at all, and unlike
\cite{zjll18,aazz19,ltlceq20}, where associations are not many-to-many). As for
simplifications to the optimization problem, our approach improves on previous
ones by considering the model's complete domain (unlike
\cite{ku16,zgh16,aazz19}, where restrictions specified by the ``patterns''
mentioned in Section~\ref{int} or similar combinatorial structures are imposed),
and by completely shunning any form of smoothing (unlike \cite{ku16,ltlceq20},
where integralities are relaxed, and unlike \cite{aav18}, where the functions
involved are approximated) and any form of problem breakup (unlike
\cite{zjll18}). Importantly, we have taken every possible precaution to make
sure the experimental setup laid down in Section~\ref{exp} is fully
reproducible. This, too, is pointedly unlike some of the previous works (most
notably the one in \cite{ku16}, whose results are hardly reproducible even at
the level of how base stations and users are deployed).

These contributions have led to the one we consider to be most important, viz.,
a clear demonstration that by looking into the model's unaltered characteristics
it is possible to glean some properties that thus far have remained unobserved
(or at least unreported). Specifically, for a reasonable set of parameter
choices the results we give in Section~\ref{res} call for a better look into the
role of macrocells in heterogeneous wireless networks.

\section{Network model}
\label{net}

We consider a set $B$ of base stations and a set $K$ of receivers. For $P_b$ the
power with which base station $b\in B$ transmits, and assuming that all
transmissions take place outdoors, the power $R_{bk}$ that reaches receiver
$k\in K$ is
\begin{equation}
R_{bk}=P_bL_b^{-1}d_{bk}^{-\tau_b},
\label{decay}
\end{equation}
where $L_b$ accounts for antenna- and frequency-related losses (as well as for
frequency- or distance-unit conversion), $d_{bk}$ is the Euclidean distance
between $b$ and $k$, and $\tau_b>2$ determines how power decays with distance.
We consider downlink communication exclusively and assume that receivers are
capable of multi-packet reception, i.e., of handling transmissions from multiple
base stations concomitantly. We model this in the manner of (uplink) CDMA over a
narrowband $N$ and a wideband $W$, therefore with processing gain $G=W/N$
\cite{tv05}.

For $B_\mathrm{on}\subseteq B$ the set of base stations currently turned on and
$b\in B_\mathrm{on}$, the SINR at receiver $k$ is
\begin{equation}
\mathrm{SINR}_{bk}=
\frac{GR_{bk}}{\gamma_0 W+\sum_{b'\in B_\mathrm{on}\setminus\{b\}}R_{b'k}},
\label{sinr}
\end{equation}
where $\gamma_0$ is the noise-floor spectral density. Decoding becomes possible
whenever
\begin{equation}
\mathrm{SINR}_{bk}\ge\beta,
\label{decode}
\end{equation}
where $\beta>1$ is a parameter related to a receiver's decoding capabilities,
assumed the same for all receivers.

If decoding is possible, we say that base station $b$ and receiver $k$ can
become \emph{associated} with each other. In this case, the maximum capacity
$C_{bk}$ for transmissions from $b$ to $k$ is given by
\begin{equation}
C_{bk}=N\log_2(1+\mathrm{SINR}_{bk}).
\label{Cbk}
\end{equation}
Moreover, the maximum number $n_\mathrm{max}$ of concomitant associations for a
given receiver is related to $G$ and $\beta$ as in
\begin{equation}
n_\mathrm{max}<1+\frac{G}{\beta},
\label{nmax}
\end{equation}
so $n_\mathrm{max}>1$ requires $\beta<G$.\footnote{See \cite{tv05}, page 142,
for a derivation of Eq.~(\ref{nmax}).}

At any given time, the total capacity provided by the network depends on how
much each of the base stations in $B_\mathrm{on}$ can provide individually. It
also depends on the current associations in the network. Clearly, any base
station $b\in B$ can always transmit as many bits per second as given by the
greatest $C_{bk}$ over any subset of $K$ (since the expression in
Eq.~(\ref{Cbk}) comes from $b$ and $k$ being associated with each other). That
is, letting $K_b\subseteq K$ be the set of receivers currently associated with
base station $b\in B_\mathrm{on}$, the transmission capacity of base station $b$
is always at least $\max_{k\in K_b}\alpha_{bk}C_{bk}$, where $\alpha_{bk}$ is
the fraction of time base station $b$ spends transmitting to receiver
$k\in K_b$. On the receivers' side, the total capacity available to receiver $k$
is $\sum_{b\in B_k}\alpha_{bk}C_{bk}$, where $B_k\subseteq B_\mathrm{on}$ is the
set of base stations currently associated with $k$. Naturally,
\begin{equation}
\sum_{k\in K_b}\alpha_{bk}\le 1
\label{alphas}
\end{equation}
and $\vert B_k\vert\le n_\mathrm{max}$ hold at all times, respectively for
every $b\in B_\mathrm{on}$ and every $k\in K$. Additionally, meeting some demand
$d_k$ at receiver $k$ requires
\begin{equation}
\sum_{b\in B_k}\alpha_{bk}C_{bk}\ge d_k.
\label{demands}
\end{equation}

\section{Mathematical formulation}
\label{opt}

The problem we address is the determination of set $B_\mathrm{on}$ and of the
fractions $\alpha_{bk}$ for every $b\in B_\mathrm{on}$ and every $k\in K$. This
is to be achieved with as little total power consumption by the base stations as
possible while satisfying the constraints given by Eqs.~(\ref{decode})
and~(\ref{nmax})--(\ref{demands}). For consistency with the goal of minimizing
power consumption, we henceforth assume that a base station is in
$B_\mathrm{on}$ if and only if at least one receiver is associated with it.

Our formulation uses two sets of variables. One of them comprises the already
seen $\alpha_{bk}$ for each $b\in B$ and $k\in K$. The other set serves to
facilitate referring to set $B_\mathrm{on}$ and to a receiver's number of
associated base stations (no greater than $n_\mathrm{max}$), as well as to the
sets $K_b$ and $B_k$ appearing respectively in Eqs.~(\ref{alphas})
and~(\ref{demands}). This second set comprises a variable $a_{bk}$ for each
$b\in B$ and $k\in K$. This variable takes its value from $\{0,1\}$, indicating
either that an association exists between base station $b$ and receiver $k$ (if
$a_{bk}=1$) or otherwise (if $a_{bk}=0$).

The $a_{bk}$'s can be combined into two useful shorthands. The first is
\begin{equation}
a_b=\max_{k\in K}a_{bk},
\end{equation}
allowing $b\in B_\mathrm{on}$ to be equated with $a_b=1$ and therefore
Eq.~(\ref{sinr}) to be rewritten as
\begin{equation}
\mathrm{SINR}_{bk}=
\frac{Ga_bR_{bk}}{\gamma_0 W+\sum_{b'\in B\setminus\{b\}}a_{b'}R_{b'k}}.
\end{equation}
The other shorthand is
\begin{equation}
\rho_b=\sum_{k\in K}a_{bk}\alpha_{bk}.
\label{sh2}
\end{equation}
Note that, while clearly $a_b=0$ implies $\rho_b=0$, the converse implication
(i.e., $a_b=1$ implies $\rho_b>0$) depends on whether $\alpha_{bk}>0$ for at
least one of the $k$'s for which $a_{bk}=1$.

The total power consumed by all base stations is given by
$P^\mathrm{sf}+P^\mathrm{tx}$, where $P^\mathrm{sf}$ refers to powering support
functions (such as cooling, signal processing, etc.) and $P^\mathrm{tx}$ refers
to powering transmission. Each base station $b$ contributes to each of these
with an amount no greater than $P_b^\mathrm{sf}$ and $P_b^\mathrm{tx}$,
respectively. Following \cite{arfb10}, a fraction $\phi_b^\mathrm{sf}$ of
$P_b^\mathrm{sf}$ is spent whenever $b$ is on ($a_b=1$), regardless of whether
it is engaged in transmissions (i.e., even if $\rho_b=0$). The complementary
fraction, $1-\phi_b^\mathrm{sf}$, is spent only when $b$ is transmitting
($\rho_b>0$). A similar split exists for $P_b^\mathrm{tx}$, now given by
fractions $\phi_b^\mathrm{tx}$ and $1-\phi_b^\mathrm{tx}$. Thus, we
have\footnote{For a logical proposition $Q$, the Iverson bracket $[Q]$ equals
$1$ if $Q$ is true, $0$ if $Q$ is false.}
\begin{align}
P^\mathrm{sf}&=\sum_{b\in B}
P_b^\mathrm{sf}
\bigl(\phi_b^\mathrm{sf}[a_b=1]+(1-\phi_b^\mathrm{sf})\rho_b\bigr),\\
P^\mathrm{tx}&=\sum_{b\in B}
P_b^\mathrm{tx}
\bigl(\phi_b^\mathrm{tx}[a_b=1]+(1-\phi_b^\mathrm{tx})\rho_b\bigr),
\end{align}
whence it follows that the transmission power $P_b$ in Eq.~(\ref{decay}) is
\begin{equation}
P_b=P_b^\mathrm{tx}(1-\phi_b^\mathrm{tx}).
\label{Pb}
\end{equation}
As a consequence, $\phi_b^\mathrm{tx}<1$ is necessary for $P_b>0$.

Given the sets $B$ and $K$ and their members' locations in Euclidean space, as
well as the values of $\gamma_0$, $\beta$, $N$, and $G$; of $L_b$, $\tau_b$,
$P_b^\mathrm{sf}$, $\phi_b^\mathrm{sf}$, $P_b^\mathrm{tx}$, and
$\phi_b^\mathrm{tx}$ for every $b\in B$; and of $d_k$ for every $k\in K$, the
optimization problem to be solved to determine all $a_{bk}$'s and all
$\alpha_{bk}$'s is the following mixed-integer nonlinear programming (MINLP)
problem.
\begin{align}
\text{minimize } &\textstyle P=P^\mathrm{sf}+P^\mathrm{tx}\label{obj}\\
\text{subject to }
&a_{bk}\in\{0,1\},&\forall b\in B, k\in K\\
&\alpha_{bk}\in[0,1],&\forall b\in B, k\in K\\
&[\mathrm{SINR}_{bk}<\beta]a_{bk}=0,&\forall b\in B, k\in K\label{c1}\\
&\textstyle\sum_{b\in B}a_{bk}<1+G/\beta,&\forall k\in K\label{c2}\\
&\textstyle\sum_{k\in K}a_{bk}\alpha_{bk}\le 1,&\forall b\in B\label{c3}\\
&\textstyle\sum_{b\in B}a_{bk}\alpha_{bk}C_{bk}\ge d_k.&\forall k\in K\label{c4}
\end{align}
Clearly, Eqs.~(\ref{c1})--(\ref{c4}) are straightforward rewrites of the
constraints given in Eqs.~(\ref{decode}) and~(\ref{nmax})--(\ref{demands}),
respectively, now making use of all the problem's variables. By Eq.~(\ref{sh2}),
the constraint in Eq.~(\ref{c3}) is equivalent to $\rho_b\le 1$. We refer to any
assignment of values to the $a_{bk}$'s and $\alpha_{bk}$'s satisfying the
constraints in Eqs.~(\ref{c1})--(\ref{c4}) as being \emph{feasible}.

\section{Experimental setup}
\label{exp}

\begin{figure}[t]
\centering
\includegraphics[scale=1.00]{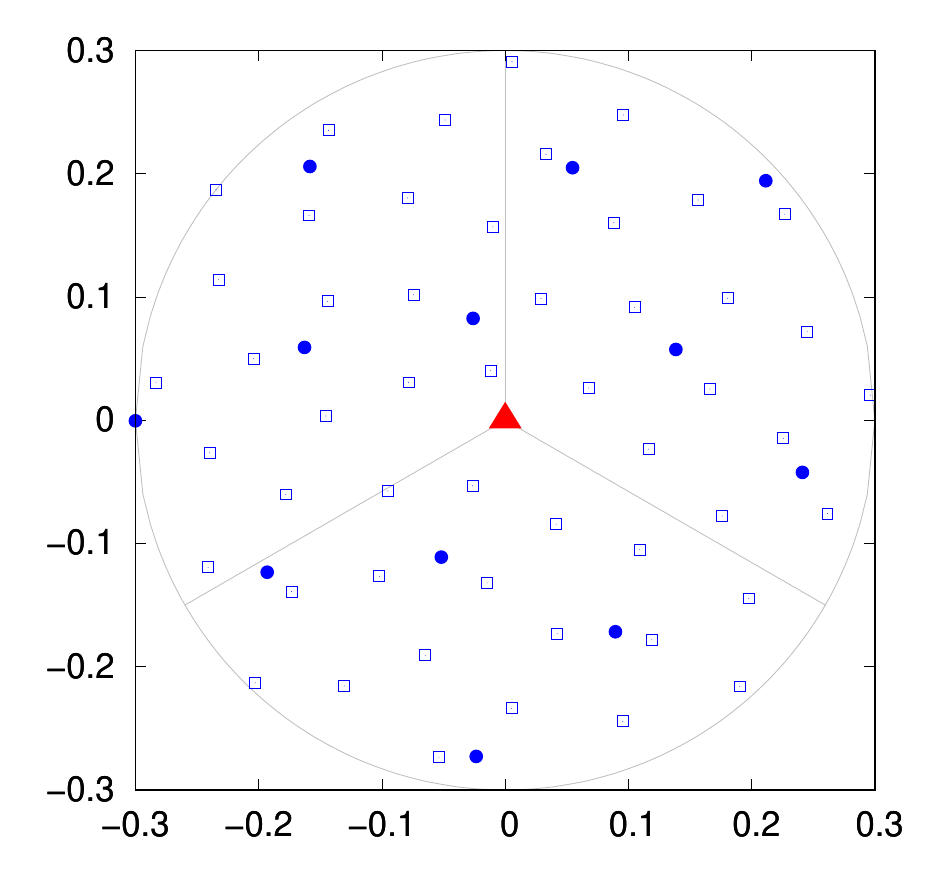}
\caption{Base-station and receiver placement for $\vert B\vert=15$,
$\vert K\vert=51$, and $R=0.3$ km. The three macrocells are represented by the
single filled triangle at the center, the twelve picocells by filled circles,
the $51$ receivers by empty squares.}
\label{fig1}
\end{figure}

We consider a two-dimensional circular region of radius $R$ in Euclidean space
and place all base stations and receivers inside this region. The set $B$ of
base stations has three macrocells and $\vert B\vert-3$ picocells. The
macrocells are placed at the circle's center, each capable of transmitting only
within an exclusive $120^\circ$-sector. The picocells can transmit in all
directions and are placed in the circle in as uniform a manner as possible. This
is achieved by mimicking the geometry of the sunflower head,\footnote{For
$n_\mathrm{loc}$ the desired number of locations, the polar coordinates $r_i$
and $\theta_i$ of the $i$th location, $1\le i\le n_\mathrm{loc}$, are
$r_i=\sqrt{i/n_\mathrm{loc}}R$ and $\theta_i=i\delta$, where
$\delta=(\sqrt{5}-1)^2\pi/2\approx 137.5^\circ$ is the golden (or Fibonacci)
angle \cite{v79}.} followed by a rotation to ensure all sectors contain the same
number of picocells (provided $\vert B\vert$ is a multiple of $3$). Receivers
are placed in the same manner as picocells.\footnote{Receivers, therefore, are
to be thought of more as test points than as users.} An illustration is given in
Figure~\ref{fig1}.

All our computational results are based on using a genetic algorithm (GA) to
solve the MINLP problem. We use brkgaAPI \cite{tr15}, an open-source,
state-of-the-art framework for efficient GA implementations, and refer to the
resulting GA as HetNetGA. The framework assumes all variables are continuous in
the interval $(0,1]$, which is consistent with the problem's $\alpha_{bk}$'s
(since these can still be arbitrarily close to $0$) but not with the $a_{bk}$'s
(since these must be either $0$ or $1$). In HetNetGA we circumvent this by
substituting a proxy $a'_{bk}\in(0,1]$ for each $a_{bk}$ and letting
$a_{bk}=[a'_{bk}>0.5]$. Moreover, all constraints must be implemented as
penalties added to the objective function. We do this by keeping a count $v$ of
constraint violations and re-expressing Eq.~(\ref{obj}) as
\begin{equation}
P=P^\mathrm{sf}+P^\mathrm{tx}+vP^\mathrm{viol},
\label{obj2}
\end{equation}
where $P^\mathrm{viol}$ is the penalty to be incurred per violation. We use
\begin{equation}
P^\mathrm{viol}=\sum_{b\in B}(P_b^\mathrm{sf}+P_b^\mathrm{tx}),
\label{pviol}
\end{equation}
that is, the maximum possible value of $P$ in Eq.~(\ref{obj}).

As a meta-heuristic, brkgaAPI can sometimes be nudged into better convergence to
feasibility and subsequent optimization by tuning its behavior to
problem-specific characteristics. We have found one such intervention to be
particularly useful when designing HetNetGA. It consists in adding a further
type of constraint to the MINLP problem in order to prevent the combined
capacity available to receiver $k$ from surpassing $d_k$ by too wide a margin.
For $\eta\in(0,1)$, the further constraint for each $k\in K$ is
\begin{equation}
\sum_{b\in B}a_{bk}(\alpha_{bk}-\eta)C_{bk}\le d_k,
\label{c5}
\end{equation}
so the capacity available to $k$, $\sum_{b\in B}a_{bk}\alpha_{bk}C_{bk}$, must
not exceed $d_k$ by more than a fraction $\eta$ of that capacity's maximum
possible value, obtained by setting every $\alpha_{bk}$ to $1$. Violating this
constraint does not alter the feasibility status of any given assignment of
values to the problem's variables but does contribute to the count $v$ affecting
Eq.~(\ref{obj2}).

Tables~\ref{tab1} and~\ref{tab2} contain all parameter values used to obtain the
results given in Section~\ref{res}. Table~\ref{tab1} refers to the formulation
of the MINLP problem, including the additional constraint in Eq.~(\ref{c5}). The
values for $L_b$ (for a center frequency of $2$ GHz and $d_{bk}$ in kilometers)
and $\tau_b$ are from \cite{3gpp}. The values for $P_b^\mathrm{sf}$,
$\phi_b^\mathrm{sf}$, $P_b^\mathrm{tx}$, and $\phi_b^\mathrm{tx}$ are loosely
based on the discussion in \cite{arfb10}. By Eq.~(\ref{Pb}), we get $P_b=39.75$
W if $b$ is a macrocell, $P_b=1$ W if $b$ is a picocell. Likewise, by
Eq.~(\ref{pviol}) we have $P^\mathrm{viol}=1\,500+(\vert B\vert-3)\,33$ W. The
values for narrowband $N$ and gain $G$ imply a wideband $W=1.28$ GHz.

Table~\ref{tab2} refers to the inner operation of brkgaAPI and its use by
HetNetGA: $p$ is the number of individuals (or chromosomes, in GA parlance) a
population has, given in proportion to the number
$n_\mathrm{var}=2\vert B\vert\vert K\vert$ of variables (or alleles);
$p_\mathrm{e}$ is the fraction of $p$ to be the elite set;
$p_\mathrm{m}$ is the fraction of $p$ to be replaced by mutants;
$\rho_\mathrm{e}$ is the probability of inheriting each allele from the elite
parent; $n_\mathrm{pop}$ is the number of independent populations; and
$n_\mathrm{gen}$ is the number of generations allowed to elapse before
termination. Each individual is an assignment of values to the problem's
variables. Owing to the strictly positive value of $p_\mathrm{e}$, HetNetGA
gives rise to an individual that minimizes $P$ in Eq.~(\ref{obj2}) globally with
as high a probability as one wishes, provided the allotted $n_\mathrm{gen}$ is
sufficiently large \cite{gm00}. Of course, we have no efficient means of
verifying whether this occurs for any given $n_\mathrm{gen}$, only of checking
individuals for feasibility.

\begin{table}[t]
\caption{Parameter values for the MINLP problem}
\label{tab1}
\centering
\renewcommand{\arraystretch}{1.25}
\input{table1}
\end{table}

\begin{table}[t]
\caption{Parameter values related to brkgaAPI}
\label{tab2}
\centering
\renewcommand{\arraystretch}{1.25}
\input{table2}
\end{table}

\section{Results, conclusion, and outlook}
\label{res}

All our results refer to the setting depicted in Figure~\ref{fig1}. Within this
setting we investigate five distinct scenarios, each forbidding certain base
stations to ever be turned on. This can be enforced for any $b\in B$ by
overriding Table~\ref{tab1} and setting $P_b^\mathrm{tx}=0$, thus leading to
$\mathrm{SINR}_{bk}=0$ for every $k\in K$, and consequently to $a_{bk}=0$ (so
$a_b=\rho_b=0$) whenever feasibility holds; cf.\ Eq.~(\ref{c1}). We refer to the
first scenario as 0m12p (no macrocells can ever be turned on, all picocells
can), and similarly for the other four scenarios: 1m12p, 2m12p, 3m12p, and 3m0p.
In relation to Figure~\ref{fig1}, the shorthand 1m refers to the upper right
sector, 2m to the two upper sectors. Notably, when no picocells are allowed to
be turned on (scenario 3m0p), by Eq.~(\ref{sinr}) it follows that
$\mathrm{SINR}_{bk}$ does not depend on the value of $G$ for any macrocell $b$
or any $k\in K$.

\begin{figure}[t]
\centering
\includegraphics[scale=0.90]{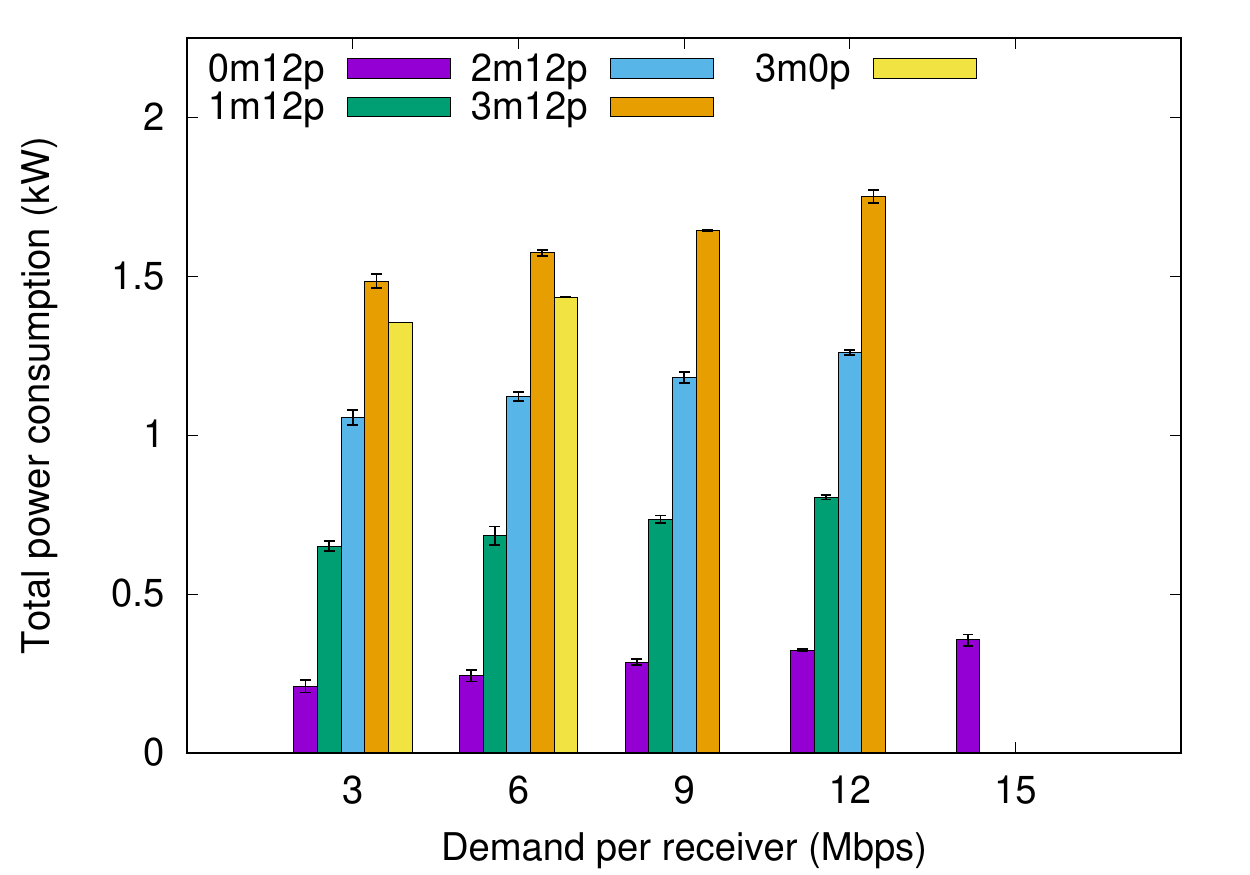}
\caption{Best value of $P$, as per Eq.~(\ref{obj}), obtained after
$n_\mathrm{gen}$ generations for each scenario and each value of $d_k$. Averages
refer to six independent runs of HetNetGA, with confidence intervals given at
the $95\%$ level. Missing values indicate that feasibility was never attained
for the corresponding combination of scenario and $d_k$ value.}
\label{fig2}
\end{figure}

\begin{figure}[t]
\centering
\includegraphics[scale=0.90]{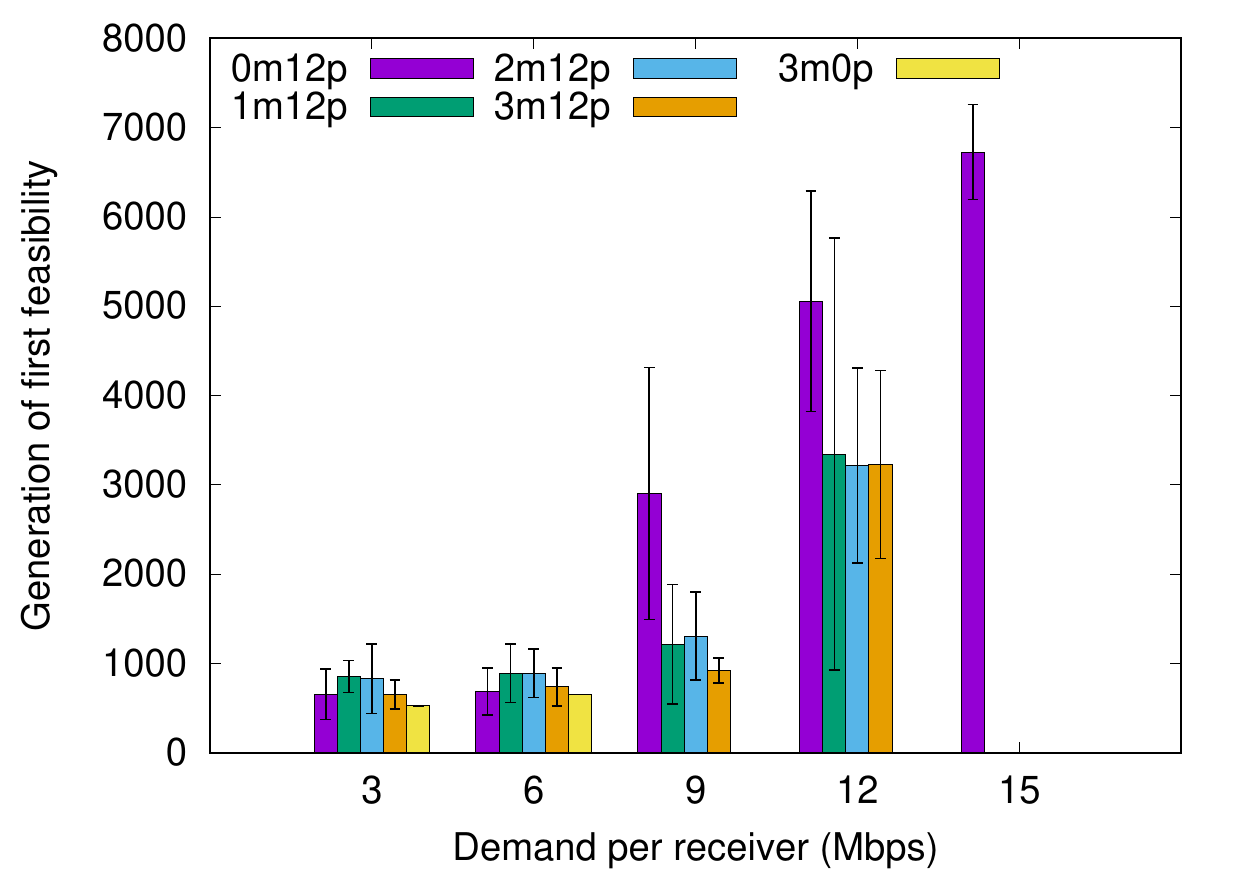}
\caption{Generation of first feasibility in the context of Figure~\ref{fig2}.
Averages refer to the same six independent runs of HetNetGA, again with
confidence intervals given at the $95\%$ level.}
\label{fig3}
\end{figure}

\begin{figure}[t]
\centering
\includegraphics[scale=1.00]{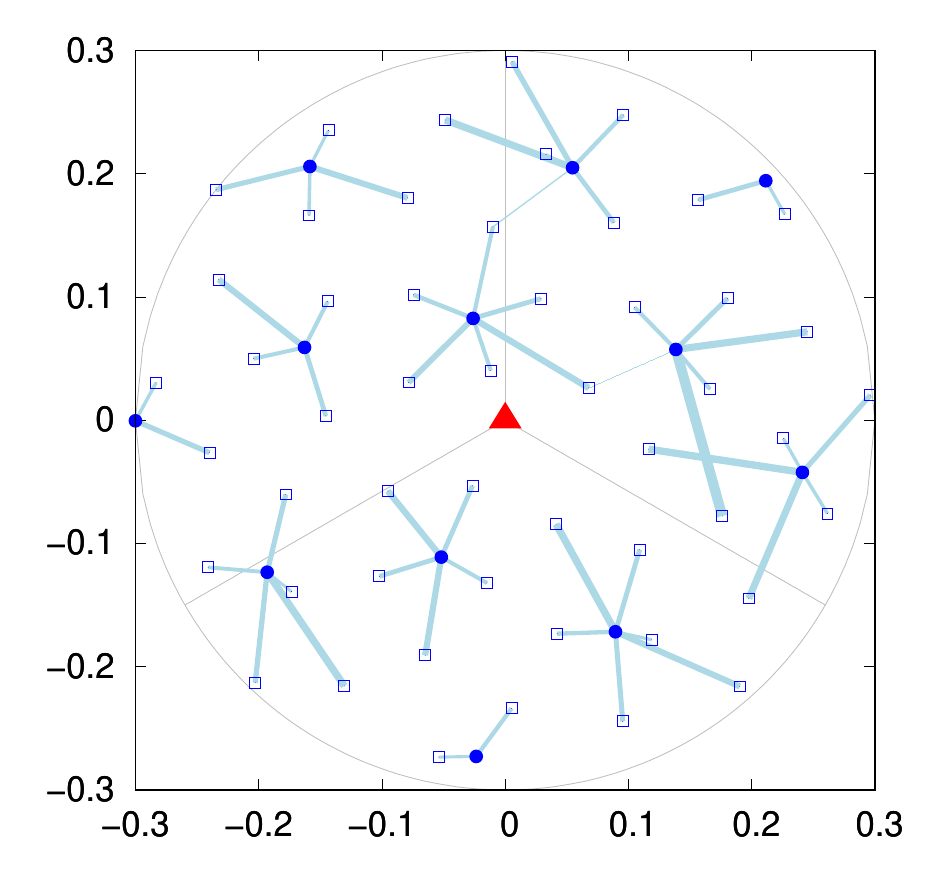}
\caption{One of the solutions contributing to Figure~\ref{fig2} ($d_k=12$ Mbps,
scenario 0m12p). Lines interconnecting receivers and picocells represent
associations. Line thickness grows linearly with the corresponding
$\alpha_{bk}$'s.}
\label{fig4}
\end{figure}

In what follows, every value we report for $P$, the total power consumed, is for
feasible individuals and follows Eq.~(\ref{obj}).\footnote{That is, having $v>0$
in Eq.~(\ref{obj2}) for such an individual implies violated constraints only of
the type given in Eq.~(\ref{c5}).} Our results are summarized in
Figures~\ref{fig2} and~\ref{fig3}, where information related to the output of
HetNetGA is given for all five scenarios and five values of demand $d_k$, the
same for every $k\in K$. For the reader's benefit, these figures are
complemented by Figure~\ref{fig4}, which relates our results to both the layout
in Figure~\ref{fig1} and the statistics in Figure~\ref{fig2}.

As expected, increasing $d_k$ increases total power consumption as well
(Figure~\ref{fig2}), and moreover makes feasibility ever harder to attain
(Figure~\ref{fig3}). Additionally, larger $\alpha_{bk}$'s tend to be necessary
as base station $b$ and receiver $k$ are placed farther apart from each other
(Figure~\ref{fig4}). Unexpectedly, though, feasibility seems to become
impossible for scenario 3m0p somewhere between $d_k=6$ and $9$ Mbps, and for the
other scenarios involving one or more macrocells (1m12p, 2m12p, and 3m12p)
somewhere between $d_k=12$ and $15$ Mbps (Figure~\ref{fig2}). This rules out the
use of macrocells for higher bit-rate demands, those for which picocells alone
will not do (this holds already for $d_k=15.5$ Mbps; data not shown). Perhaps
some sweet spot exists at which this happens, but locating it has proven
elusive. As far as we have been able to observe, any assistance a macrocell
might provide in meeting a certain bit-rate demand is offset by the interference
it causes, and then the whole setting becomes energetically disadvantageous.

Naturally, the flip side of this conclusion is that HetNetGA, in spite of its
properties of convergence to a global minimum, is after all the one to blame.
Some support against this possibility is given in Figure~\ref{fig3}, which shows
how early feasibility is first attained during the $n_\mathrm{gen}$ generations.
This happens ever later as $d_k$ increases, and also with confidence intervals
much greater than those of Figure~\ref{fig2}. The suggestion here is that,
notwithstanding all the variation in the number of generations to hit first
feasibility, by the $n_\mathrm{gen}$th generation the solutions HetNetGA outputs
are approximately equivalent to one another. This hardly rules out the
abovementioned sweet spot, but does make it hard enough to find to cast doubt on
the practicality of looking for it.

Thus, insofar as the model outlined in Section~\ref{net} can be said to describe
the system under study faithfully, a role is yet to be found for macrocells.
Further research should concentrate on variations of this model, and also of its
characteristics as summarized in Table~\ref{tab1}, aiming to better delimit what
can be expected of macrocells. HetNetGA, which inherently preserves the
mathematical description of the associated MINLP problem, is expected to remain
a useful tool.

\subsection*{Acknowledgments}

This work was supported in part by Conselho Nacional de Desenvolvimento
Cient\'\i fico e Tecnol\'ogico (CNPq), Coordena\c c\~ao de Aperfei\c coamento de
Pessoal de N\'\i vel Superior (CAPES), and a BBP grant from Funda\c c\~ao Carlos
Chagas Filho de Amparo \`a Pesquisa do Estado do Rio de Janeiro (FAPERJ).

\bibliography{hetnet}
\bibliographystyle{unsrt}

\end{document}

%% file: table1.tex
\begin{tabular}{ccc}
\hline
\multicolumn{1}{c}{Macrocells} & \multicolumn{1}{c}{Picocells} & \multicolumn{1}{c}{Decoding} \\
\hline
$L_b=128.1$ dBm & $L_b=140.7$ dBm & $\gamma_0=1.174\times 10^{-20}$ J \\
$\tau_b=3.76$ & $\tau_b=3.67$ & $\beta=5$ dB \\
$P_b^\mathrm{sf}=235$ W & $P_b^\mathrm{sf}=28$ W & $N=10$ MHz \\
$\phi_b^\mathrm{sf}=0.85$ & $\phi_b^\mathrm{sf}=0.5$ & $G=128$ \\
\cline{3-3}
$P_b^\mathrm{tx}=265$ W & $P_b^\mathrm{tx}=5$ W & Eq.~(\ref{c5}) \\
\cline{3-3}
$\phi_b^\mathrm{tx}=0.85$ & $\phi_b^\mathrm{tx}=0.8$ & $\eta=0.005$ \\
\hline
\end{tabular}

%% file: table2.tex
\begin{tabular}{ccc}
\hline
$p=10n_\mathrm{var}$ & $p_\mathrm{e}=0.2$ & $p_\mathrm{m}=0.1$ \\
$\rho_\mathrm{e}=0.4$ & $n_\mathrm{pop}=3$ & $n_\mathrm{gen}=10\,000$ \\
\hline
\end{tabular}